\documentclass[aps,prb,twocolumn,superscriptaddress]{revtex4}

\usepackage{graphicx}
\usepackage{amssymb}
\usepackage[dvipdfm,colorlinks,citecolor=blue,linkcolor=blue,urlcolor=black]{hyperref}

\begin{document}
%\linenumbers

\title{High-resolution angle-resolved photoemission spectroscopy study of the electronic structure of EuFe$_2$As$_2$}

\author{Bo Zhou}
\author{Yan Zhang}
\author{Le-Xian Yang}
\author{Min Xu}
\author{Cheng He}
\author{Fei Chen}
\author{Jia-Feng~Zhao}
\author{Hong-Wei Ou}
\author{Jia Wei}
\author{Bin-Ping Xie}
\affiliation{State Key Laboratory of Surface Physics, Department of Physics, and Advanced Materials Laboratory, Fudan University, Shanghai 200433, People's Republic of China}

\author{Tao Wu}
\author{Gang Wu}
\affiliation{Hefei National Laboratory for Physical Sciences at Microscale and Department of Physics, University of Science and Technology of China, Hefei, Anhui 230026, People's Republic of China}

\author{Masashi Arita}
\author{Kenya~Shimada}
\author{Hirofumi Namatame}
\author{Masaki Taniguchi}
\affiliation{Hiroshima Synchrotron Radiation Center and Graduate School of Science, Hiroshima University, Hiroshima 739-8526, Japan}

\author{X. H. Chen}
\affiliation{Hefei National Laboratory for Physical Sciences at Microscale and Department of Physics, University of Science and Technology of China, Hefei, Anhui 230026, People's Republic of China}

\author{D. L. Feng}
\email[]{dlfeng@fudan.edu.cn}
\affiliation{State Key Laboratory of Surface Physics, Department of Physics, and Advanced Materials Laboratory, Fudan University, Shanghai 200433, People's Republic of China}

\date{\today}

\begin{abstract}

We report the high-resolution angle-resolved photoemission spectroscopy studies of electronic structure of EuFe$_2$As$_2$. The paramagnetic state data are found to be consistent with density-functional calculations. In the antiferromagnetic ordering state of Fe, our results show that the band splitting, folding, and hybridization evolve with temperature, which cannot be explained by a simple folding picture. Detailed measurements reveal that a tiny electron Fermi pocket and a tiny hole pocket are formed near $(\pi,\pi)$ in the $(0,0)$-$(\pi,\pi)$ direction, which qualitatively agree with the results of quantum oscillations, considering $k_z$ variation in Fermi surface. Furthermore, no noticeable change within the energy resolution is observed across the antiferromagnetic transition of Eu$^{2+}$ ordering, suggesting weak coupling between Eu sublattice and FeAs sublattice.

\end{abstract}

%\keywords{}

\maketitle

\section{Introduction}

The discovery of superconductivity with transition temperatures up to the record of 56~K in iron-based superconductors has generated intensive research on this new class of high-temperature superconductors.\cite{Kamihara,Chen_Sm1,ZARen_Sm,Chen_Sm2,Wang_NdTh} The  parent compounds of these superconductors often exhibit a ground state of the spin density wave (SDW) ordering of the Fe moments.\cite{Rotter_Ba} Like the cuprates, upon proper doping, superconductivity emerges while the magnetic order is suppressed.\cite{Rotter_BaK,Chen_BaK,Sasmal_SrK,Chen_CaNa,Qi_EuNa,Jeevan_EuK,Anupam_EuK} In  the underdoped regime of certain iron pnictides, superconductivity may even coexist with SDW,\cite{Chen_coexist,YZhang} which once again highlights the intimate relation between superconductivity and magnetism in such unconventional superconductors.

Numerous studies have been devoted to elucidate the nature of the SDW in iron pnictides. In angle-resolved photoemission spectroscopy (ARPES) studies of BaFe$_2$As$_2$ and SrFe$_2$As$_2$, exotic band splittings were observed,\cite{LXYang,YZhang,GDLiu,ZX_SDW}  which were suggested to be related with the local moments, and responsible for the SDW transition.\cite{LXYang,YZhang,GDLiu} However, some other studies, including ARPES, optical measurements, and quantum oscillation measurements, support the conventional Fermi surface nesting mechanism of SDW.\cite{Hasan_Sr,NLWang_SDW,Sebastian,Analytis} Alternatively, a picture that involves both local and itinerant elements was suggested.\cite{Moon_optic,Medici,Mazin} So far, a full understanding of the SDW in iron pnictides is still missing.

\begin{figure}[b]
\includegraphics[width=6cm]{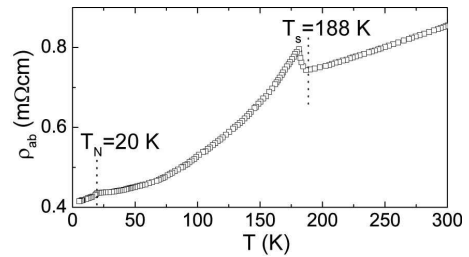}
\caption{\label{fig1} Temperature dependence of in-plane resistivity for EuFe$_2$As$_2$ single crystal. (Ref.~\onlinecite{Chen_EuLa})}
\end{figure}

EuFe$_2$As$_2$ is special in the so-called ``122'' series ($A$Fe$_2$As$_2$, with $A$=Ba, Ca, Sr, or Eu, etc.) of iron pnictides, because it contains large local moments of Eu$^{2+}$ ions ($\sim$6.8~$\mu_B$) on $A$ site.\cite{YXiao_Eu} Aside from the SDW/structural transition at $T_S$=188~K, another transition at $T_N$=20~K has been found which is associated with an $A$ type antiferromagnetic ordering of Eu$^{2+}$ ions (see Fig.~\ref{fig1}).\cite{Raffius1993,ZARen_Eu,Chen_EuLa,YXiao_Eu,Herrero_Eu}

To reach a comprehensive picture of the complex electronic structure in the SDW state of iron pnictides, and reveal the  manifestation of various magnetic orderings on the  electronic structure, we have performed the ARPES measurements of EuFe$_2$As$_2$ single crystals.  Two hole Fermi pockets around $(0,0)$ and two electron pockets around  $(\pi,\pi)$ are observed in the high-temperature paramagnetic (PM) state. In the SDW state, band splitting, folding, and hybridization are found to  evolve with  temperature. Detailed measurements reveal that the two small Fermi pockets symmetric about $(\pi,\pi)$ in the $(0,0)$-$(\pi,\pi)$ direction are electronlike and holelike, respectively.  This asymmetric electronic structure has not yet been reported in ARPES studies of iron pnictides, but is consistent with the quantum oscillation results by considering the different $k_z$s for the two small pockets.\cite{Sebastian,Analytis} The drastic changes in electronic structure, i.e., the band splitting and the band shift, cannot be explained by a simple folding picture. In agreement with several studies by different techniques,\cite{YXiao_Eu,Herrero_Eu,Eu_optic} no noticeable change in ARPES measurement is found across the antiferromagnetic transition of Eu$^{2+}$ ordering.

\begin{figure}
\includegraphics[width=8.5cm]{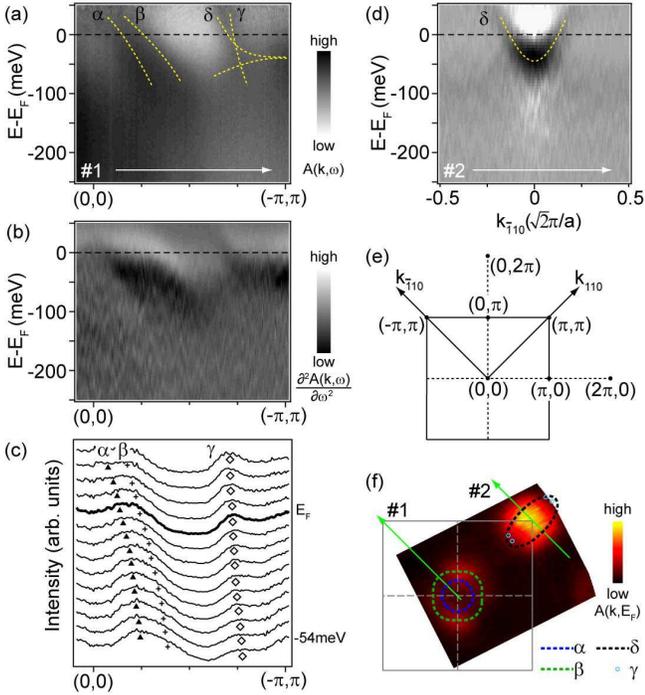}
\caption{\label{fig2}(Color online)  Electronic structure in the paramagnetic state measured at 195~K. (a) Photoemission intensities divided by the energy-resolution-convoluted Fermi-Dirac function to reveal band dispersions in the vicinity of $E_F$, along cut 1 [$(0,0)$-$(\pi,\pi)$] as indicated in panel f, and the corresponding (b) second derivative image plot with respect to energy, and (c) MDCs. (d) The second derivative image plot along cut 2 as indicated in panel f. Dashed lines (in panels a and d) and markers (in panel c) are eye guides of the band structure. In panel a, bands except for $\gamma$ are determined from the second derivative data in panel b. (e) Schematics of the 2D paramagnetic BZ, with notations of high symmetry points and axes. (f) The Fermi surface map with an integration window of 10~meV about $E_F$, overlaid by Fermi surface contours. The $\gamma$ Fermi crossings are shown by circles, wherever identifiable. See text for details.}
\end{figure}

\section{Experiment}

High quality EuFe$_2$As$_2$ single crystals  were synthesized by self-flux method, and more details can be found in Ref.~\onlinecite{Chen_EuLa}. Its stoichiometry was confirmed by energy dispersive x-ray (EDX) analysis. ARPES measurements were performed with randomly polarized 21.2~eV photons from a helium discharge lamp and with circularly polarized synchrotron light from Beamline 9 of Hiroshima synchrotron radiation center (HSRC). Scienta R4000 electron analyzers are equipped in both setups. The overall energy resolution is 9~meV, and angular resolution is 0.3$^\circ$. The samples were cleaved \textit{in situ}, and measured in ultrahigh vacuum below 4$\times10^{-11}$~mbar. All data reported here were taken within 4~h after cleaving to minimize the aging effect. The high quality sample surface is confirmed by a clear pattern of low-energy electron diffraction (LEED). Most data presented here were taken with the helium discharge lamp unless otherwise specified.

\section{Results and discussions}

\begin{figure}
\includegraphics[width=8.5cm]{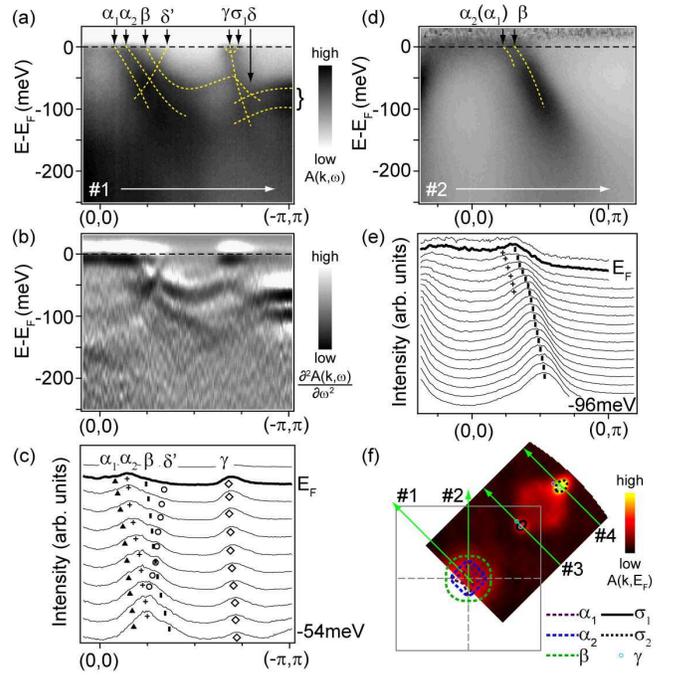}
\caption{\label{fig3}(Color online) Electronic structure in the SDW state measured at 22~K. (a) Photoemission intensities of cut 1 [along $(0,0)$-$(\pi,\pi)$] as indicated in panel f, and the corresponding (b) second derivative image plot, and (c) MDCs. (d) Photoemission intensities of cut 2 [along $(0,0)$-$(0,\pi)$] as indicated in panel f, which are individually normalized by the integrated weight of the according MDC. (e) The corresponding MDCs for panel d. Dashed lines (in panels a and d) and markers (in panels c and e) are eye guides of the band structure. In panel a, bands except for $\gamma$ are determined from the second derivative data in panel b. Bands in panel d are determined from MDCs in panel e. (f) The SDW state Fermi surface map measured at 22~K with an integration window of 10~meV about $E_F$, overlaid by Fermi surface contours. Similar to Fig.~\ref{fig2}(f), the $\gamma$ Fermi crossings are shown by circles, wherever identifiable. See text for details.}
\end{figure}

\begin{figure*}
\includegraphics[width=16cm]{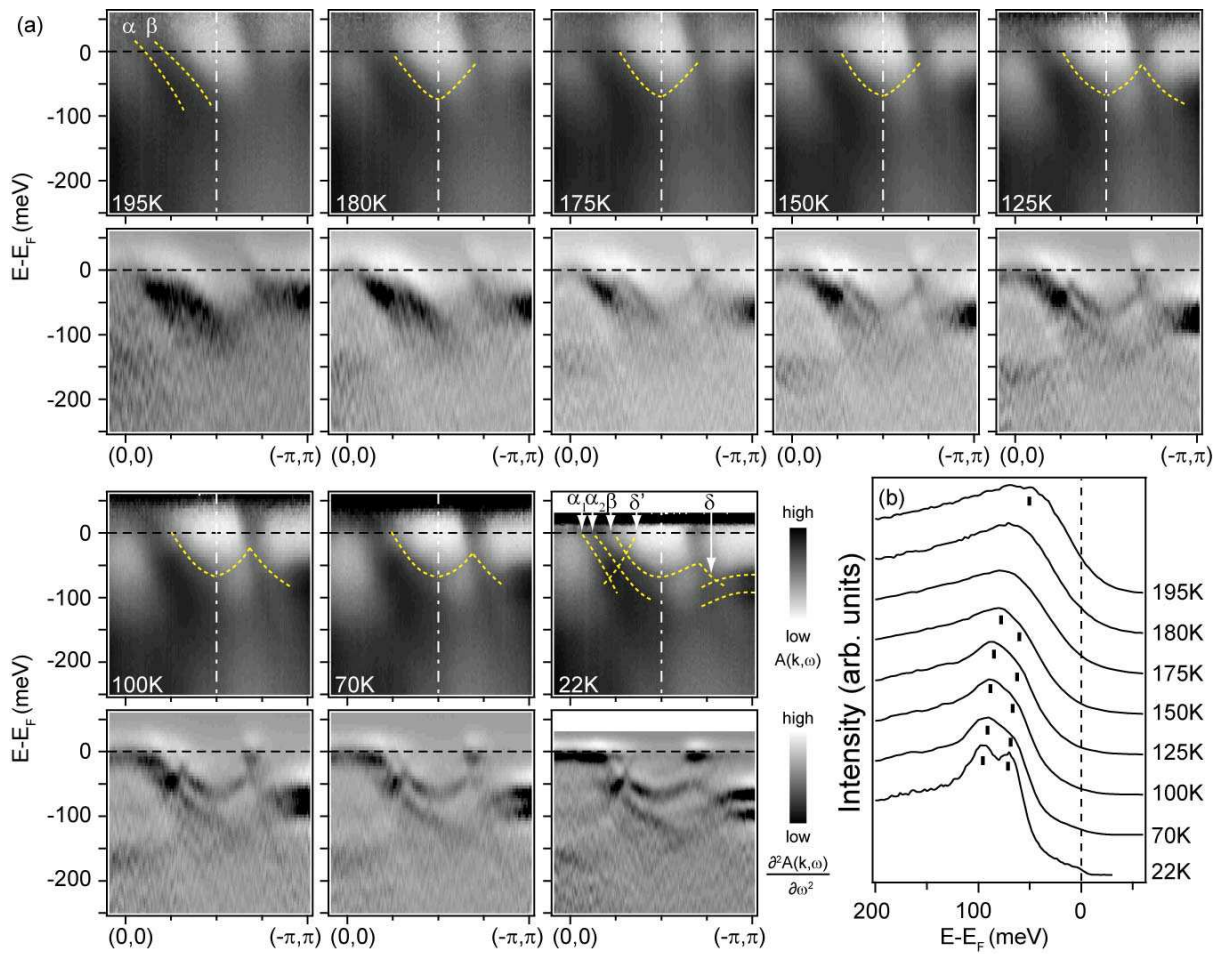}
\caption{\label{fig4}(Color online) The temperature dependence of band structure along $(0,0)$-$(\pi,\pi)$ of EuFe$_2$As$_2$. (a) Photoemission intensities divided by the energy-resolution-convoluted Fermi-Dirac function and the second derivative image plots at 195, 180, 175, 150, 125, 100, 70, and 22~K, respectively. Dashed lines are eye guides for selected bands, determined from the second derivative data. (b) Temperature dependence of a single EDC at the $(\pi,\pi)$ point. Markers are eye guides to the peak positions.}
\end{figure*}

\begin{figure}
\includegraphics[width=8.5cm]{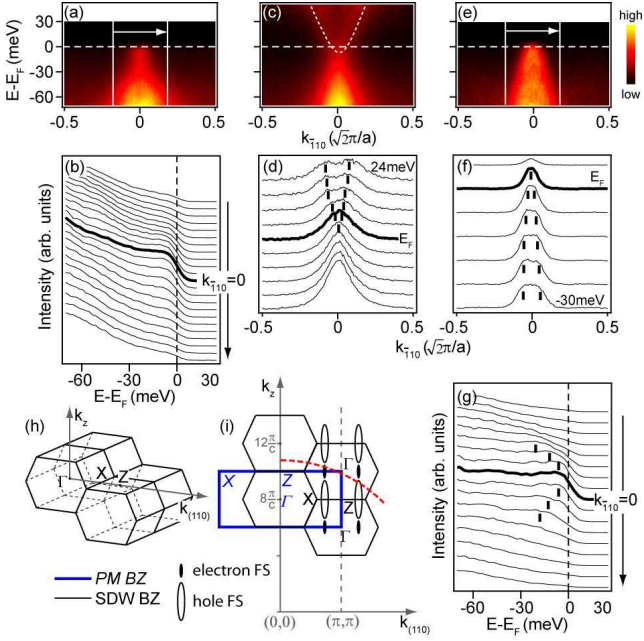}
\caption{\label{fig5}(Color online) (a) Photoemission intensities of cut 3 [indicated in Fig.~\ref{fig3}(f)] measured at 22~K and (b) the corresponding EDCs in the range as indicated in panel a. (c) Its photoemission intensities measured at 175~K divided by the energy-resolution-convoluted Fermi-Dirac function. The dashed line is the guide to eyes of an electron band. (d) The corresponding MDCs for panel c. (e) Photoemission intensities of cut 4 [indicated in Fig.~\ref{fig3}(f)] measured at 22~K and the corresponding (f) MDCs, and (g) EDCs in the range as indicated in panel e. The markers are guides to eyes to trace the dispersion of the electronlike or holelike band. (h) 3D SDW folded BZ. (i) PM and SDW BZs in $k_{110}k_z$ plane as well as an illustration of small electron and hole Fermi surfaces. The red dashed curve is a constant energy cut representing 21.2~eV [$k_z$=10.8~$\pi$/c at $(0,0)$]. Black characters denote high symmetry points of 3D SDW BZ and blue italic characters denote for paramagnetic BZ.}
\end{figure}

\begin{figure*}
\includegraphics[width=17cm]{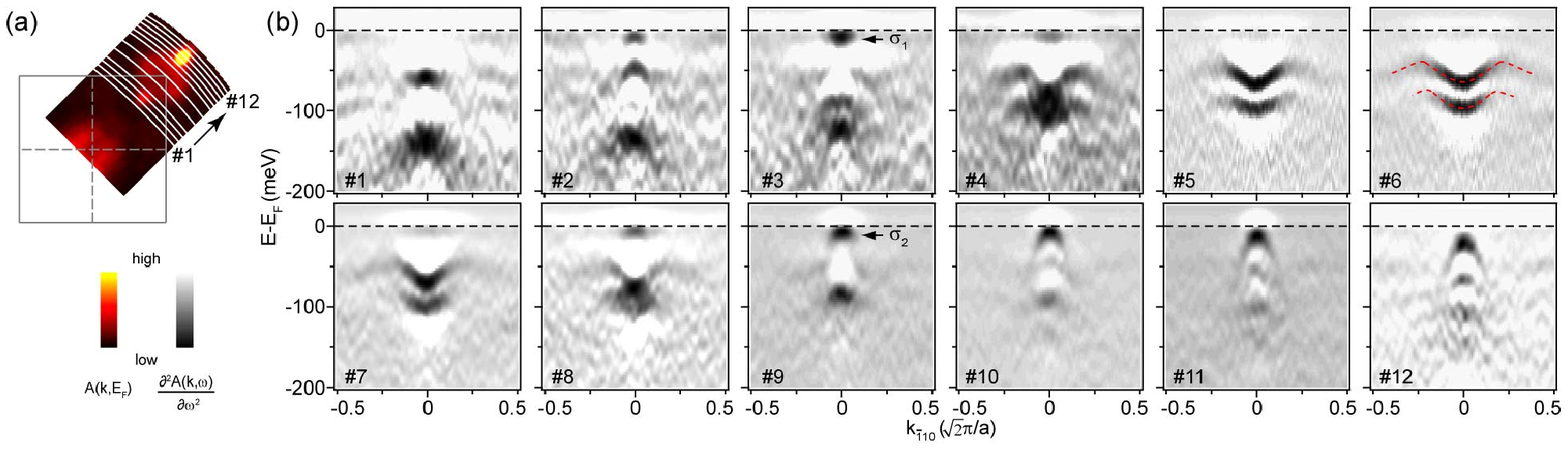}
\caption{\label{fig6}(Color online) Electronic structure around  $(\pi,\pi)$  in SDW state. (a) The SDW state Fermi surface map  and indications of  cuts 1-12. (b) The second derivative image plots along cuts 1-12.}
\end{figure*}

The electronic structure in the paramagnetic state is shown in Fig.~\ref{fig2}, which is similar to other 122 compounds reported by ARPES measurements.\cite{ZX_PM,Hasan_Sr} Figures~\ref{fig2}(a)-\ref{fig2}(c) show the data of cut 1 [along $(0,0)$-$(\pi,\pi)$ as indicated in Fig.~\ref{fig2}(f)], where two holelike bands ($\alpha$ and $\beta$) can be identified. From the raw data [Fig.~\ref{fig2}(a)] and the corresponding momentum distribution curves (MDCs) [Fig.~\ref{fig2}(c)], an electronlike band ($\gamma$) can be identified. Although the band bottom of $\gamma$ is not resolved in the data, it may reach about 200~meV below the Fermi energy ($E_F$) by following its energy dispersion. In the second derivative data with respect to energy [Fig.~\ref{fig2}(b)], the feature of $\gamma$ is not resolved  due to the large Fermi velocity. Nonetheless, there is another weak electronlike feature ($\delta$) near $(\pi,\pi)$ with smaller Fermi velocity and with band bottom at about 40-50~meV. Along $(2\pi,0)$-$(\pi,\pi)$-$(0,2\pi)$ [cut 2 in Fig.~\ref{fig2}(d)], a parabolic electronlike band is observed with the band bottom at 45~meV below $E_F$, which should be assigned to $\delta$ according to the position of the band bottom. Notice that photoemission intensity of the $\delta$ band is much less than that of $\gamma$ along $(0,0)$-$(\pi,\pi)$, but this is reversed away from $(0,0)$-$(\pi,\pi)$, most likely caused by matrix element effects.\cite{Hasan_FeTe} The resulting Fermi surfaces are depicted in Fig.~\ref{fig2}(f), which consist of two hole pockets around $(0,0)$, an elongated elliptical electron pocket around $(\pi,\pi)$, and some Fermi crossings of $\gamma$ around $(\pi,\pi)$. Here the $\gamma$ Fermi crossings may be originated from the second electron pocket which has been reported before in BaFe$_2$As$_2$ and SrFe$_2$As$_2$,\cite{ZX_PM,Hasan_Sr} although its complete shape cannot be determined from our data, likely due to matrix element effects. The measured Fermi surface topology, especially around $(\pi,\pi)$, is in agreement with theoretical calculations for 122 compounds,\cite{Singh_calc,Ma_calc,Nekrasov_calc} since the measured $(0,0)$-$(\pi,\pi)$ direction in the two-dimensional (2D) BZ is closer to the $Z$-$X$ direction in the three-dimensional (3D) paramagnetic BZ under 21.2~eV light [Fig.~\ref{fig5}(i)].

Figure~\ref{fig3} shows the electronic structure in the SDW state. Dispersions of various bands are shown as dashed lines in Figs.~\ref{fig3}(a), \ref{fig3}(b), and \ref{fig3}(d). Comparing with the paramagnetic band structure, the $\gamma$ band does not change along $(0,0)$-$(\pi,\pi)$. However, $\alpha$ splits into $\alpha_1$ and $\alpha_2$ in the SDW state. In addition, the $\delta$ band hybridizes with $\beta$, and a folded band ($\delta^\prime$) of $\delta$ appears around $(0,0)$. Moreover, the feature around $(\pi,\pi)$ below $E_F$  in the paramagnetic state splits into two inverted parabolic bands in the SDW state [as denoted by the bracket in Fig.~\ref{fig3}(a)], which are likely to be the dispersions from the split $\alpha_1$ and $\alpha_2$ bands. Along $(0,0)$-$(0,\pi)$, $\alpha_1$ and $\alpha_2$ are nearly degenerate, and folding is not resolved in this direction.  $\alpha_1$, $\alpha_2$, $\beta$, and $\delta^\prime$ around $(0,0)$, and $\gamma$ around $(\pi,\pi)$ are identified to cross $E_F$ by examining the MDCs in Figs.~\ref{fig3}(c) and \ref{fig3}(e). As a result, the Fermi surface around $(0,0)$ consists of three hole Fermi pockets and a folded electron Fermi pocket. Again similar to the paramagnetic state, $\gamma$ is not observable away from $(0,0)$-$(\pi,\pi)$ due to the possible matrix element effects. As observed in the SDW state Fermi surface [Fig.~\ref{fig3}(f)], a distinctive difference from the paramagnetic state Fermi surface is that two ``bright spots" appear in the $(0,0)$-$(\pi,\pi)$ direction near  $(\pi,\pi)$ . As shown in Fig.~\ref{fig3}(a), the hybridization of $\delta$ and $\beta$ results in a possible small electron band whose bottom is barely below $E_F$, and gives a ``bright spot'' (the $\sigma_1$ pocket) in the photoemission intensity map of the first BZ, which shall be given a zoom-in in Fig.~\ref{fig5}. Thereby, no gap is found at $E_F$ in the electronic structure. Note that the $\sigma$ bands are a result of the hybridization between the $\delta$ and $\beta$ bands. However, the $\gamma$ band appears to cross the $\sigma$ bands without any sign of hybridization. This is most likely due to their different symmetries with respect to certain mirror plane, as illustrated in Ref.~\onlinecite{YZhang_orbital} for the BaFe$_{1.85}$Co$_{0.15}$As$_2$ in the paramagnetic state, where the $\gamma$ band shows up only  in $\pi$-polarized geometry while the $\delta$ and $\beta$ bands  show up only in the $\sigma$-polarized geometry. Therefore the $\delta$ and $\beta$ bands are both of odd symmetry, and thus the   $\sigma$ band from their hybridization may also be odd. On the other hand, the $\gamma$ band is even. The opposite symmetries of $\gamma$ and $\sigma$ could result in the observed absence of hybridization here, assuming the symmetry properties of the bands are preserved in  this series of iron pnictides.

The evolution of electronic structure with temperature is illustrated in Fig.~\ref{fig4}. As the temperature decreases from $T_S$, the $\beta$ band appears to be symmetric with respect to the midway of $(0,0)$-$(-\pi,\pi)$ (the BZ boundary of the SDW state) due to the magnetic ordering in the SDW state. Nevertheless, the $\beta$ band becomes \emph{asymmetric} as the temperature is further lowered to 22~K, since this band hybridizes with $\delta$ and therefore does not cross $E_F$. This hybridization strengthens with decreasing temperature [Fig.~\ref{fig4}(a)]. On the other hand, the $\beta$ band crosses the folded band $\delta^\prime$ with much weaker hybridization near the $(0,0)$ point. As observed before in other 122 systems,\cite{LXYang,YZhang,GDLiu,ZX_SDW} the band splitting is observed in this material as well. From the paramagnetic state to SDW state, the $\alpha$ band near $(0,0)$ and the inverted parabolic band near $(\pi,\pi)$ split into $\alpha_1$, $\alpha_2$ and two inverted parabolic bands correspondingly. The remarkable temperature dependence of the two inverted parabolic bands is shown in Fig.~\ref{fig4}(b), where single energy distribution curves (EDCs) at the $(\pi,\pi)$ point are stacked. At 22~K, the two peaks are at 70 and 95~meV below $E_F$, respectively, which correspond to the band tops of the two inverted parabolic bands. Both bands move toward $E_F$ with increasing temperature, and merge into one broad feature in the paramagnetic state. When two features are distinguishable, their separation decreases slightly with increasing temperature.  If considering a simple folding picture as suggested in Ref.~\onlinecite{Hasan_Sr}, it cannot reproduce the complex band structure in the SDW state, especially inadequate to explain the splitting  and shift of bands.\cite{ZX_SDW} Therefore, local exchange interactions are suggested to be included to explain the electronic structure.\cite{LXYang,YZhang}

To further investigate the nature of the ``bright spots", a cut perpendicular to $(0,0)$-$(\pi,\pi)$ and across the $\sigma_1$ pocket is measured as indicated by cut 3 in Fig.~\ref{fig3}(f). In data taken at  22~K [Figs.~\ref{fig5}(a) and \ref{fig5}(b)],  a hump is observed near $E_F$ around $k_{\overline{1}10}$=0, indicating most probably a small electron pocket. To further elucidate this, additional data were taken at higher temperature (175~K) but still in the SDW state and shown in Figs.~\ref{fig5}(c) and \ref{fig5}(d). Figure~\ref{fig5}(c) shows the photoemission intensity divided by the energy-resolution-convoluted Fermi-Dirac function. It reveals a parabolic electron band in the vicinity of $E_F$. The corresponding MDCs are shown in Fig.~\ref{fig5}(d), where markers indicate the dispersion of the electron band. The separated two peaks above $E_F$ merge into a single peak slightly below $E_F$, indicating that this electron band indeed crosses $E_F$.

Similarly, data were taken along a cut through the ``bright spot'' near $(\pi,\pi)$ outside the first BZ [cut 4 indicated in Fig.~\ref{fig3}(f)]. Interestingly, a dispersion that gives a hole Fermi pocket is observed in the spectrum image plot, EDCs, and MDCs [Figs.~\ref{fig5}(e)-\ref{fig5}(g)]. These results indicate that the two small pockets symmetric about $(\pi,\pi)$ are electronlike and holelike respectively. This asymmetric electronic structure can be understood by realizing the $k_z$ variation in the Fermi surface. Figure~\ref{fig5}(h) presents a 3D SDW folded BZ, and Fig.~\ref{fig5}(i) shows the corresponding SDW BZ together with the paramagnetic BZ in the $k_{110}k_z$ plane. Along the in-plane $k_{110}$ direction, Fermi surfaces are depicted in Fig.~\ref{fig5}(i) that pairs of small electron and hole pockets are symmetric about $\Gamma$ and $Z$, respectively, whereas a dashed curve is a schematic constant energy cut through an electron pocket and a hole pocket near $(\pi,\pi)$. This scenario naturally explains our result, and is in qualitative agreement with quantum oscillations of 122 compounds, which indicate the existence of small electron and hole Fermi pockets aligned along $k_z$.\cite{Sebastian,Analytis}

The estimated Luttinger volume of the $\alpha_1$, $\alpha_2$, and $\beta$ Fermi pockets near $(0,0)$ are 1.7\%, 3.6\%, and 9.5\% of the paramagnetic BZ, respectively. The tiny $\sigma_1$ and $\sigma_2$ pockets near $(\pi,\pi)$ both comprise less than 0.5\% of the paramagnetic BZ. These results agree well with the quantum oscillation measurements for the small Fermi pockets ($\alpha_1$, $\sigma_1$, and $\sigma_2$). However, quantum oscillations did not find the large Fermi surfaces ($\alpha_2$ and $\beta$). This might be due to the fact that the electrons could not finish  circulating the large pocket before they are scattered by  impurities.

\begin{figure}
\includegraphics[width=7cm]{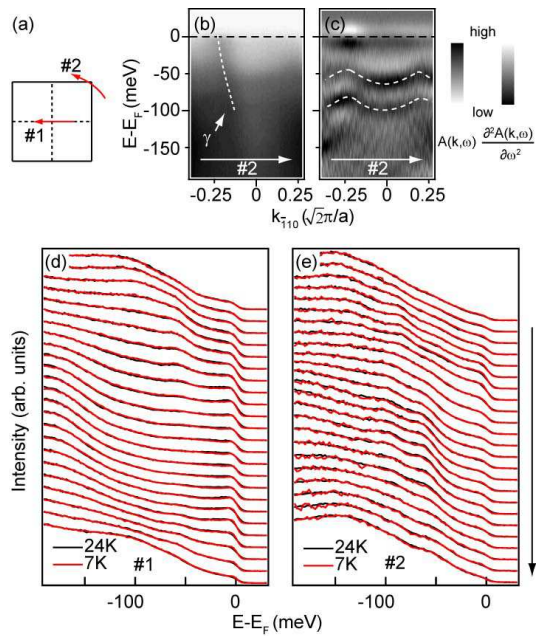}
\caption{\label{fig7}(Color online) (a) Cuts 1 and 2 around $(0,0)$  and $(\pi,\pi)$  are indicated in the BZ, respectively. (b) Photoemission intensities along cut 2 (indicated in panel a) measured at 7~K, and (c) the corresponding second derivative image plot. (d) and (e) The comparison of EDCs at 7~K and 24~K, along cuts 1 and 2 (indicated in panel a), respectively. Here data were taken with the synchrotron light of 10~eV in HSRC.}
\end{figure}

To understand the detailed electronic structure evolution near $(\pi,\pi)$, a series of the second derivative image plots are presented in Fig.~\ref{fig6}(b). Away from cut 6, the two $M$-like features move toward the higher energies and become blurred. Features at $E_F$ are only visible near the two ``bright spots''. Although the Fermi surface mapping shows patch-like features near $(\pi,\pi)$  other than the two ``bright spots'' [Fig.~\ref{fig6}(a)], the patch is just remnant spectral weight from bands below $E_F$. As shown in cuts 5 and 5-7 of Fig.~\ref{fig6}(b), the maxima of two $M$-like features are below $E_F$, therefore these features do not cross $E_F$. Similar to the observations  made in BaFe$_2$As$_2$,\cite{ZX_SDW,GDLiu} it is observed that the electronic structure shows saddle-surface-like features near $(\pi,\pi)$ below $E_F$, as seen in Fig.~\ref{fig3}(a) and cut 6 in Fig.~\ref{fig6}(b).  Band tops of the two inverted parabolic bands coincide with band bottoms of the two parabolic bands, therefore they are the same bands split from the paramagnetic bands. Unlike the presence of the $\gamma$ band along $(0,0)$-$(\pi,\pi)$ in Fig.~\ref{fig3}(a), it is absent along the direction perpendicular to $(0,0)$-$(\pi,\pi)$ [cuts 5-7 in Fig.~\ref{fig6}(b)]. However, it is observed in a similar direction measured with 10~eV synchrotron light [Fig.~\ref{fig7}(b)]. Therefore it is indeed a manifestation of matrix element effects that little photoemission intensity of $\gamma$ is detected away from $(0,0)$-$(\pi,\pi)$ under the experimental setup of helium lamp.

Since the Eu$^{2+}$ ions in EuFe$_2$As$_2$ undergo an antiferromagnetic transition at a Neel temperature of $T_N$=20~K, it is thus straightforward to investigate the electronic structure below and above this transition.  Figures~\ref{fig7}(b) and \ref{fig7}(c) present the photoemission intensities and the corresponding second derivative around $(\pi,\pi)$ taken at 7~K with 10~eV photons. The $\gamma$ band and two $M$-like bands are clearly observed. The band bottoms of the $M$-like bands are situated at the same energies as those measured with 21.2~eV photons. Since the ARPES spectra are much more bulk sensitive when measured with 10~eV photons, the match between 21.2~eV data and 10~eV data suggests that the shift and splitting of the two parabolic (or inverted) bands near $(\pi,\pi)$ are not just surface effect, but reflecting bulk properties. Figures~\ref{fig7}(d) and \ref{fig7}(e) show the EDCs around $(0,0)$ and $(\pi,\pi)$, respectively, both below and above $T_N$. Across $T_N$, no observable change except the thermal broadening has been observed within the accuracy allowed by the energy resolution of 8~meV. In Ref.~\onlinecite{Chen_EuLa}, it was found that the $A$-type antiferromagnetism can be converted to ferromagnetism with a magnetic field as small as 1.5~T and the inter-layer coupling was estimated to be 0.46-0.69~meV, which is one order of magnitude lower than our resolution. This shows that the electronic structure near $E_F$ is not altered noticeably by the ordering of the Eu$^{2+}$ moments; on the other hand, even if there were certain subtle effects of the magnetism, they cannot be easily detected in our ARPES measurements, as limited by the current energy resolution. Nonetheless, magnetism or fluctuations may play an important role to mediate superconductivity, noting that Eu$_{1-x}$La$_x$Fe$_2$As$_2$ is not superconducting,\cite{Chen_EuLa} while Eu$_{1-x}$K$_x$Fe$_2$As$_2$ is superconducting where long-range magnetic ordering is killed.\cite{Jeevan_EuK}

\section{Conclusion}

In conclusion, we have presented in detail the electronic structure of EuFe$_2$As$_2$ both in the paramagnetic state, the SDW state, and the antiferromagnetically ordered state of the Eu$^{2+}$ moments. The temperature evolution of the band dispersion across $T_S$ reveals that the band folding and splitting are established in the SDW state. The folded band $\beta$ hybridizes with  $\delta$ and the hybridization gap opens as the temperature decreases. In addition, this hybridization gives rise to a small electron band near $(\pi,\pi)$  in the first BZ. However, the similar small pocket outside the first BZ is proved to be holelike. This gives a natural explanation of the observation of quantum oscillation measurements. A simple folding picture cannot describe the significant change in the electronic structure in the SDW state. The experiment by more bulk sensitive ARPES with 10~eV light provides evidence that the band splitting is not due to any surface effect, such as a surface reconstruction. The undetectable change in electronic structure across $T_N$ suggests the weak electronic coupling between Eu sublattice and FeAs sublattice.

During the preparation of this paper, we noticed that another ARPES study of EuFe$_2$As$_2$ has been posted online.\cite{ARPES_Eu} Most experimental results of both papers are consistent with each other, although the determined Fermi surfaces are not exactly the same, probably due to different $k_z$s.

\begin{acknowledgments}

Some of the preliminary data (not shown here) were taken at ELETTRA synchrotron light source. This work was supported by the NSFC, MOE, MOST (National Basic Research Program No. 2006CB921300), STCSM of China, and the NSF of U.S. under Grant No. PHY-0603759, and ICTP-ELETTRA project.

\end{acknowledgments}


\begin{references}

\bibitem{Kamihara} Y. Kamihara, T. Watanabe, M. Hirano, and H. Hosono, J. Am. Chem. Soc. \textbf{130}, 3296 (2008).

\bibitem{Chen_Sm1} X. H. Chen, T. Wu, G. Wu, R. H. Liu, H. Chen, and D. F. Fang, Nature (London) \textbf{453}, 761 (2008).

\bibitem{ZARen_Sm} Z. A. Ren, W. Lu, J. Yang, W. Yi, X.-L. Shen, Z. Cai, G.-C. Che, X.-L. Dong, L.-L. Sun, F. Zhou, and Z.-X. Zhao, Chin. Phys. Lett. \textbf{25}, 2215 (2008).


\bibitem{Chen_Sm2} R. H. Liu, G. Wu, T. Wu, D. F. Fang, H. Chen, S. Y. Li, K. Liu, Y. L. Xie, X. F. Wang, R. L. Yang, L. Ding, C. He,   D. L. Feng, and X. H. Chen, Phys. Rev. Lett. \textbf{101}, 087001 (2008).


\bibitem{Wang_NdTh} C. Wang, L.-J. Li, S. Chi, Z.-W. Zhu, Z. Ren, Y.-K. Li, Y.-T. Wang, X. Lin, Y.-K. Luo, S. Jiang, X.-F. Xu, G.-H. Cao, and Z. A. Xu, EPL \textbf{83}, 67006 (2008).


\bibitem{Rotter_Ba} M. Rotter, M. Tegel, D. Johrendt, I. Schellenberg, W. Hermes, and R. P\"ottgen, Phys. Rev. B \textbf{78}, 020503(R) (2008).

\bibitem{Rotter_BaK} M. Rotter, M. Tegel, and D. Johrendt, Phys. Rev. Lett. \textbf{101}, 107006 (2008).


\bibitem{Chen_BaK} G. Wu, R. H. Liu, H. Chen, Y. J. Yan, T. Wu, Y. L. Xie, J. J. Ying, X. F. Wang, D. F. Fang, and X. H. Chen, EPL \textbf{84}, 27010 (2008).

\bibitem{Sasmal_SrK} K. Sasmal, B. Lv, B. Lorenz, A. M. Guloy, F. Chen, Y.-Y. Xue, and C.-W. Chu, Phys. Rev. Lett. \textbf{101}, 107007 (2008).

\bibitem{Chen_CaNa} G. Wu, H. Chen, T. Wu, Y. L. Xie, Y. J. Yan, R. H. Liu, X. F. Wang, J. J. Ying, and X. H. Chen, J. Phys.: Condens. Matter \textbf{20}, 422201 (2008).

\bibitem{Qi_EuNa} Y.-P. Qi, Z.-S. Gao, L. Wang, D.-L. Wang, X.-P. Zhang, and Y.-W. Ma, New. J. Phys. \textbf{10}, 123003 (2008).

\bibitem{Jeevan_EuK} H. S. Jeevan, Z. Hossain, D. Kasinathan, H. Rosner, C. Geibel, and P. Gegenwart, Phys. Rev. B \textbf{78}, 092406 (2008).

\bibitem{Anupam_EuK} Anupam, P. L. Paulose, H. S. Jeevan, C. Geibel, and Z. Hossain, J. Phys.: Condens. Matter \textbf{21}, 265701 (2009).

\bibitem{Chen_coexist} H. Chen, Y. Ren, Y. Qiu, W. Bao, R. H. Liu, G. Wu, T. Wu, Y. L. Xie, X. F. Wang, Q. Huang, and X. H. Chen, EPL \textbf{85}, 17006 (2009).

\bibitem{YZhang} Y. Zhang, J. Wei, H. W. Ou, J. F. Zhao, B. Zhou, F. Chen, M. Xu, C. He, G. Wu, H. Chen, M. Arita, K. Shimada, H. Namatame, M. Taniguchi, X. H. Chen, and D. L. Feng, Phys. Rev. Lett. \textbf{102}, 127003 (2009).

\bibitem{LXYang} L. X. Yang, Y. Zhang, H. W. Ou, J. F. Zhao, D. W. Shen, B. Zhou, J. Wei, F. Chen, M. Xu, C. He, Y. Chen, Z. D. Wang,  X. F. Wang, T. Wu, G. Wu, X. H. Chen, M. Arita, K. Shimada, M. Taniguchi, Z. Y. Lu, T. Xiang, and D. L. Feng, Phys. Rev. Lett. \textbf{102}, 107002 (2009).

\bibitem{GDLiu} G.-D. Liu, H.-Y. Liu, L. Zhao, W.-T. Zhang, X.-W. Jia, J.-Q. Meng, X.-L. Dong, J. Zhang, G. F. Chen, G.-L. Wang, Y. Zhou, Y. Zhu, X.-Y. Wang, Z.-Y. Xu, C.-T. Chen, and X. J. Zhou, Phys. Rev. B \textbf{80}, 134519 (2009).

\bibitem{ZX_SDW} M. Yi, D. H. Lu, J. G. Analytis, J.-H. Chu, S.-K. Mo, R.-H. He, M. Hashimoto, R. G. Moore, I. I. Mazin, D. J. Singh,  Z. Hussain, I. R. Fisher, and Z.-X. Shen, Phys. Rev. B \textbf{80}, 174510 (2009).

\bibitem{Hasan_Sr} D. Hsieh, Y. Xia, L. Wray, D. Qian, K. K. Gomes, A. Yazdani, G. F. Chen, J. L. Luo, N. L. Wang, and M. Z. Hasan, arXiv:0812.2289 (unpublished).

\bibitem{NLWang_SDW} W. Z. Hu, J. Dong, G. Li, Z. Li, P. Zheng, G. F. Chen, J. L. Luo, and N. L. Wang, Phys. Rev. Lett. \textbf{101}, 257005 (2008).

\bibitem{Sebastian} S. E. Sebastian, J. Gillett, N. Harrison, P. H. C. Lau, D. J. Singh, C. H. Mielke, and G. G. Lonzarich, J. Phys.: Condens. Matter \textbf{20}, 422203 (2008).

\bibitem{Analytis} J. G. Analytis, R. D. McDonald, J.-H. Chu, S. C. Riggs, A. F. Bangura, C. Kucharczyk, M. Johannes, and I. R. Fisher, Phys. Rev. B \textbf{80}, 064507 (2009).

\bibitem{Moon_optic} S. J. Moon, J. H. Shin, D. Parker, W. S. Choi, I. I. Mazin, Y. S. Lee, J. Y. Kim, N. H. Sung, B. K. Cho, S. H. Khim, J. S. Kim, K. H. Kim, and T. W. Noh, arXiv:0909.3352 (unpublished).

\bibitem{Medici} L. de¡¯ Medici, S. R. Hassan, and M. Capone, J. Supercond. Novel Magn. \textbf{22}, 535 (2009).

\bibitem{Mazin} M. D. Johannes and I. I. Mazin, Phys. Rev. B \textbf{79}, 220510(R) (2009).

\bibitem{YXiao_Eu} Y. Xiao, Y. Su, M. Meven, R. Mittal, C. M. N. Kumar, T. Chatterji, S. Price, J. Persson, N. Kumar, S. K. Dhar, A. Thamizhavel, and Th. Brueckel, Phys. Rev. B \textbf{80}, 174424 (2009).

\bibitem{Chen_EuLa} T. Wu, G. Wu, H. Chen, Y. L. Xie, R. H. Liu, X. F. Wang, and X. H. Chen, J. Magn. Magn. Mater. \textbf{321}, 3870 (2009).

\bibitem{Raffius1993} H. Raffius, E. M\"orsen, B. D. Mosel, W. M\"uller-Warmuth, M. Jeitschko, L. Terb\"uchte, and T. Vomhof, J. Phys. Chem. Solids \textbf{54}, 135 (1993).

\bibitem{ZARen_Eu} Z. A. Ren, Z.-W. Zhu, S. Jiang, X.-F. Xu, Q. Tao, C. Wang, C.-M. Feng, G. H. Cao, and Z. A. Xu, Phys. Rev. B \textbf{78}, 052501 (2008).


\bibitem{Herrero_Eu} J. Herrero-Mart\'in, V. Scagnoli, C. Mazzoli, Y.-X. Su, R. Mittal, Y.-G. Xiao, T. Brueckel, N. Kumar, S. K. Dhar, A. Thamizhavel, and L. Paolasini, Phys. Rev. B \textbf{80}, 134411 (2009).

\bibitem{Eu_optic} D. Wu, N. Bari\ifmmode \check{s}\else \v{s}\fi{}i\ifmmode \acute{c}\else \'{c}\fi{}, N. Drichko, S. Kaiser, A. Faridian, M. Dressel, S. Jiang, Z. Ren, L. J. Li, G. H. Cao, Z. A. Xu, H. S. Jeevan, and P. Gegenwart, Phys. Rev. B \textbf{79}, 155103 (2009).

\bibitem{ZX_PM} M. Yi, D. H. Lu, J. G. Analytis, J.-H. Chu, S.-K. Mo, R.-H. He, R. G. Moore, X. J. Zhou, G. F. Chen, J. L. Luo, N. L. Wang, Z. Hussain, D. J. Singh, I. R. Fisher, and Z.-X. Shen, Phys. Rev. B \textbf{80}, 024515 (2009).

\bibitem{Hasan_FeTe} Y. Xia, D. Qian, L. Wray, D. Hsieh, G. F. Chen, J. L. Luo, N. L. Wang, and M. Z. Hasan, Phys. Rev. Lett. \textbf{103}, 037002 (2009).


\bibitem{Singh_calc} D. J. Singh, Phys. Rev. B \textbf{78}, 094511 (2008).

\bibitem{Ma_calc} F.-J. Ma, Z.-Y. Lu, and T. Xiang, Front. Phys. China. (to be published).


\bibitem{Nekrasov_calc} I. A. Nekrasov, Z. V. Pchelkina, and M. V. Sadovskii, Pis'ma Zh. Eksp. Teor. Fiz. \textbf{88}, 155 (2008) [JETP Lett. \textbf{88}, 144 (2008)].

\bibitem{YZhang_orbital} Y. Zhang, B. Zhou, F. Chen, J. Wei, M. Xu, L. X. Yang, C. Fang, W. F. Tsai, G. H. Cao, Z. A. Xu, M. Arita, H. Hayashi, J. Jiang, H. Iwasawa, C. H. Hong, K. Shimada, H. Namatame, M. Taniguchi, J. P. Hu, D. L. Feng, arXiv:0904.4022v2 (unpublished).

\bibitem{ARPES_Eu} S. de Jong, E. van Heumen, S. Thirupathaiah, R. Huisman, F. Massee, J. B. Goedkoop, R. Ovsyannikov, J. Fink, H. A. D\"urr, A. Gloskovskii, H. S. Jeevan, P. Gegenwart, A. Erb, L. Patthey, M. Shi, R. Follath, A. Varykhalov, and M. S. Golden, EPL \textbf{89}, 27007 (2010).


\end{references}
\end{document}